\documentclass[useAMS,usenatbib,usedcolumn]{mn2e} 
\usepackage{epsfig}	
\usepackage{times}	
\title[Bayesian modelling of N4325 ]{Bayesian modelling of the cool core galaxy group NGC 4325}
\author[Russell, Ponman $\&$ Sanderson]
       {Paul A. Russell \thanks{E-mail: par@star.sr.bham.ac.uk},
        Trevor J. Ponman and Alastair J. R. Sanderson\\
 School of Physics and Astronomy, University of Birmingham, Edgbaston, Birmingham B15 2TT, UK \\
       \\}
 \date{Accepted 2007 February 22.}

\pagerange{\pageref{firstpage}--\pageref{lastpage}}
\pubyear{2007}

\shortcites{}

\begin{document}

\maketitle

\label{firstpage}

\begin{abstract}
  
 \noindent

We present an X-ray analysis of the radio-quiet cool-core galaxy group NGC 4325 (z=0.026) based on Chandra and ROSAT observations. The Chandra data were analysed using XSPEC deprojection, 2D spectral mapping and forward-fitting with parametric models. Additionally, a Markov chain Monte Carlo method was used to perform a joint Bayesian analysis of the Chandra and ROSAT data. The results of the various analysis methods are compared, particularly those obtained by forward-fitting and deprojection. The spectral mapping reveals the presence of cool gas displaced up to 10 kpc from the group centre. The Chandra X-ray surface brightness shows the group core to be highly disturbed, and indicates the presence of two small X-ray cavities within 15 kpc of the group core. The XSPEC deprojection analysis shows that the group has a particularly steep entropy profile, suggesting that an AGN outburst may be about to occur. With the evidence of prior AGN activity, but with no radio emission currently observed, we suggest that the group in in a pre-outburst state, with the cavities and displaced gas providing evidence of a previous, weak AGN outburst. 

\end{abstract}


\section{Introduction}
\label{sec:intro}

Many of the most massive galaxy clusters which have been observed have central gas cooling times which are shorter than the Hubble time. If no additional heat sources are present, the gas is expected to cool and condense at the centres of these clusters. The systems which are likely candidates for this scenario have been called `cooling-flow' clusters. However, despite the large number of observed massive clusters which have suitably short cooling times, none have been found which exhibit significant amounts of gas cooling below X-ray temperatures \citep[e.g.][]{tam01, kaa04}. This suggests that there must be sources of additional heating which are able to prevent cooling flows from occurring, and the erstwhile cooling-flows are now commonly referred to as `cool cores'. Such cores are cooler at their centres, typically by a factor of $\leq3$.

Besides the lack of cool gas \citep[e.g.][]{edg03} or stars which ought to have been formed by cooling flows, additional evidence for non-gravitational heating has come from the study of cluster scaling relations. Without any significant non-gravitational heating or radiative cooling, the entropy of galaxy clusters should scale with the mean cluster temperature. This self-similar scaling has been found to be inconsistent with observations \citep[e.g.][]{pon03, pra06}, which instead suggests that entropy scales approximately as $T^{2/3}$. 

On the basis that one or more forms of non-gravitational gas heating would best account for the observational evidence, the viability of various heating mechanisms has been explored in recent years. Many of these mechanisms have been invoked in an effort to find an effective method of heat transfer from hot outer gas to the cooler core, for example by thermal conduction or dynamical friction.  Unfortunately, most of the proposed heating methods appear unable to provide the amounts of energy required to prevent cooling flows \citep[e.g.][]{kim03, hab04, kim05}.

Perhaps the most successful candidate for gas heating in cluster cores has been feedback from active galactic nuclei (AGN). In this scenario, the intracluster medium (ICM) is periodically heated by outbursts from a central AGN, and this released energy offsets at least some of the radiative cooling of the ICM. The precise way in which the energy released by the AGN heats the the ICM is still not well understood, although hydrodynamical simulations suggest that such energy releases may form bubbles of low density gas which rise through the ICM \citep[e.g.][]{qui01}. This process can also reduce the cooling time by transporting gas to regions of lower pressure \citep[e.g.][]{chu02, dal04}.

There is some observational evidence to support the AGN hypothesis for cool core clusters. A majority of these clusters contain active radio sources, some of which exhibit emission coincident with cavities in the ICM \citep[e.g.][]{mcn00, bir04}. These observations provide direct evidence of the interaction between outflows from AGN and the surrounding ICM.

It is important to note, however, that not all cool core clusters contain an active radio source. \citet{don05} observed two radio quiet clusters (A1650 and A2244) with Chandra, which are classified as cool core systems. They found no evidence of fossil X-ray cavities which would indicate previous AGN heating events. They also found that the cluster cooling times were significantly longer ($\sim$1Gyr) than that of cooling flow clusters with active radio sources (~100Myr). They conclude that either the central gas has not yet cooled to a point which would trigger AGN feedback, or that the gas underwent a `major heating event' more than 1Gyr ago, and has not required any feedback since that time to prevent catastrophic cooling.

Clearly radio-quiet cool core clusters provide an important test of the AGN heating hypothesis. By examining such clusters for evidence of previous AGN activity, it may be possible to verify whether radio-quiet clusters have undergone AGN heating, if some other mechanism is required to explain any lack of central cooling, or indeed whether they might be cooling more rapidly than other clusters. Galaxy groups are particularly good candidates in which to look for such evidence. This is because groups have shorter cooling times than clusters, which means that feedback mechanisms, if present, are likely to occur more frequently in groups than in clusters. In addition, because galaxy groups are smaller than clusters, the effects of any feedback processes are likely to be more easily observable in groups.

In this paper we perform a joint Chandra and ROSAT analysis of the radio-quiet cool core group, NGC 4325, looking at the radial structure of the gas properties and searching for evidence of previous AGN activity. The analysis of this group provides the opportunity to compare the results of different analysis techniques, specifically a forward-fitting technique and the PROJCT XSPEC deprojection model. The Chandra and ROSAT observations, and the data reduction, are outlined in Section~\ref{sec:chandra_reduction}. The analysis techniques employed are presented in Section~\ref{sec:analysis}. The results are given in Section~\ref{sec:results}, with a discussion and conclusions in Sections~\ref{sec:discuss} and \ref{sec:conclusion} respectively.

\begin{figure}
\centering
\includegraphics[width=7.8cm, angle=0]{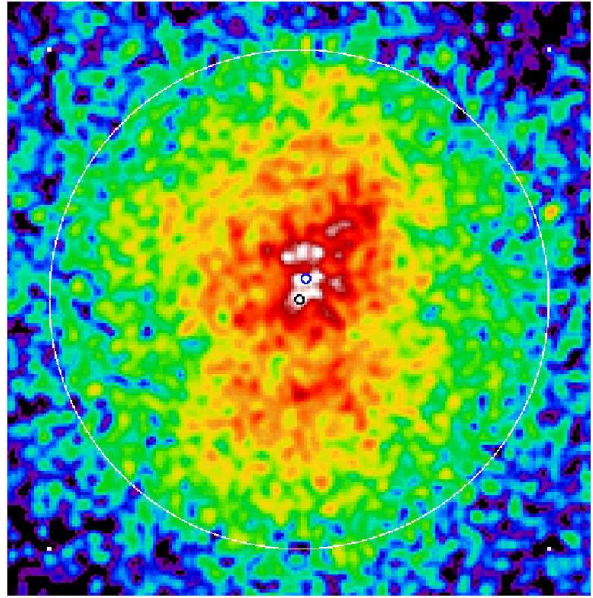}
\caption{Smoothed Chandra X-ray image of the N4325 group core. The core of the group exhibits clear asymmetric structure, despite apparent symmetry beyond 20 kpc from the group centre, indicated by the white circle. The brightest X-ray emission appears aligned (northwest-southeast) with the cool gas visible in the spectral temperature map (Fig.~\ref{spectemp}). The north-south axis visible in the gas at r $\sim$10--20 kpc corresponds to the orientation of the central elliptical galaxy (Fig.~\ref{optical}). The two circles in the centre of the image indicate the centroid of the group gas (lower circle) and optical centre of the elliptical galaxy (upper circle), which are separated by approximately 2 kpc.\label{smooth}}
\end{figure}

\section{Chandra Data Reduction}
\label{sec:chandra_reduction}

The NCG~4325 galaxy group was observed using Chandra (Obsid: 3232) on February 4, 2003. The 30.46 ks exposure was produced using the Advanced CCD Imaging Spectrometer (ACIS-S) in VFAINT mode. In order to filter flares from the data, two lightcurves were generated, one for CCD 7 and another for CCDs 2,3,6 and 8. The lightcurves were filtered using the Markevitch lc$\_$clean script, which resulted in the removal of 5.3 ks, or 17.6 per cent of the original data. Events due to bad pixels and afterglows were also removed. The time-dependent gain variation of the instrument was compensated for using corr$\_$tgain, a component of the CIAO Chandra data analysis software package.

Background subtraction was performed by selecting an appropriate Markevitch blank-sky dataset (http://cxc.harvard.edu/contrib/maxim/bg/). In order to allow for small variations in the particle background between the blank-sky field and the target observation, the normalisation of the background dataset was rescaled according to the ratio of count rates in the particle-dominated 7--12 keV energy range. This ratio was calculated separately for each chip in the NGC~4325 dataset, to check for consistent results. In order to exclude the flux from contaminating sources from 
this ratio, such features were identified and excluded using the procedure outlined in \citet{san05}, which we summarise as follows.

Images were extracted in the 0.5--2.0 keV band each CCD chip in both the target and blank sky datasets. The main image was then searched for sources with the ciao task WAVDETECT using a background image created by smoothing
the blank sky image with a Gaussian of width 1 arcmin, to suppress Poisson noise. The source regions identified in this way were masked out of both the main and blank sky datasets, and the remaining counts in the 7--12 keV band were summed. A rescale factor was then determined as the ratio of the net count rate in the background dataset to that in the main dataset. The effective exposure time of the blank sky dataset was then multiplied by this rescale factor and the process repeated, in order to adjust the blank-sky counts, until no new source regions were found. The final rescale factor applied to the blank sky background was taken as the mean of the final rescale factors calculated for the individual CCD chips, which was 0.98.

Finally, to exclude contaminating point sources from the subsequent analysis of emission from NGC~4325, a similar source detection scheme was adopted, using WAVDETECT. This time, however, a background image was constructed by lightly smoothing the NGC~4325 image (with a Gaussian of $\sigma=1.5$ pixels), and then median filtering the resulting image with a sliding box of width 32 pixels (using the FTOOLS task `fmedian'). Each point source was then associated with a Ciao region file, for simple removal from the cleaned image. A total of 13 point sources were found using this method. The initial light smoothing stage is necessary to reduce the Poisson noise, without smearing small scale features too much, to allow the median filter to sample the characteristic average flux on spatial scales larger than those of such point sources.

The advantage of this approach is that small scale features are removed, while larger regions of extended emission (corresponding to hot gas of interest in the analysis) are left intact (but with Poisson fluctuations smoothed out). This provides a more effective means of estimating a background for the purposes of identifying point sources than simply using the blank sky field. This is because Poisson fluctuations in brighter regions of extended emission will appear to be significant compared to the blank sky field, and would thus be mistakenly identified as contaminating sources.

\section{Data Analysis}
\label{sec:analysis}

\subsection{XSPEC Deprojection}
\label{sec:analysis_xspec}

The Chandra data were prepared as described in Section~\ref{sec:chandra_reduction}. The regions used for XSPEC analysis were defined by circular annuli, centred on the group core at CCD coordinates x=4160, y=3803. In order to determine the number of  annuli, they were defined such that each annulus contained a minimum of 3000 net counts per bin, using all counts within a 0.5-0.7 keV energy range. Using these criteria, eleven annuli were generated.

The spectra for each annulus were generated using the CIAO dmextract command on each of the previously defined annuli. Additional keywords, required by the XSPEC PROJCT model were applied to the fits headers for each of the generated spectra. These keywords, XFLT0001, XFLT0002, and XFLT0003 define the semi-major axis, semi-minor axis and position angle of each annulus respectively. As every annulus was circular for this analysis, each of these values was set to zero. Weighted rmf and arf files were generated to correct for variations in the detector response over the extracted regions. This was done by taking the average of contributions from areas of the detector which have different responses and weighting them relative to the distribution of counts in the 0.5-–2.0 keV band.

The extracted spectra were fitted using the PROJCT model in XSPEC 11.3.1. The PROJCT model performs a 3D to 2D projection of prolate ellipsoidal shells onto elliptical annuli, using a standard `onion peeling' technique. The PROJCT model does this by calculating the fraction of each ellipsoidal volume which is intersected by the cross section of each annulus. By starting at the outer annulus and working inwards, it is then possible to subtract the contributions to each annulus from the annuli which lie beyond it. This of course depends on the assumption that the outermost annulus agregates all the remaining emission from outside this region.

The PROJCT model was combined with a standard two component model, composed of a single temperature MEKAL plasma model and a galactic absorption component (WABS). The fitting was performed with the redshift fixed at z = 0.0257 and with an energy range between 0.5 and 7.0 keV, using Andres \& Grevesse abundances. The free parameters during fitting were the MEKAL temperature, density normalisation and abundance, in addition to the hydrogen column density.

\begin{figure}
\centering
\includegraphics[width=8.4cm, angle=0]{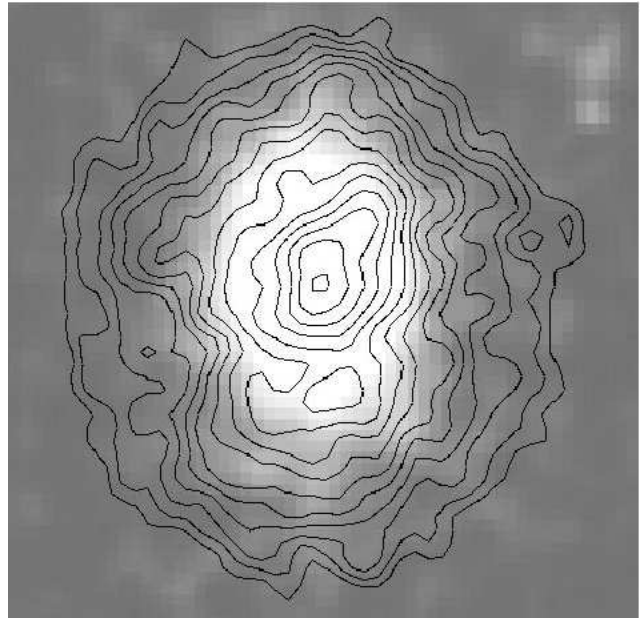}
\caption{Smoothed optical DSS image with overlaid X-ray contours extending to approximately 20 kpc. Note the asymmetrical X-ray contours in the centre of the image, in contrast to the relatively featureless central elliptical galaxy.\label{optical}}
\end{figure}

\subsection{Cluster Fitting}
\label{sec:analysis_fitting}

The cluster-fitting software used in this analysis was developed at the University of Birmingham, designed originally to enable reduction and analysis of ROSAT and ASCA X-ray data. The fitting software uses routines from the Asterix X-ray analysis package \citep{eyl91, llo00}. A major feature of the cluster-fitting software is its ability to perform spatial and spectral fitting of user-specified models to constrain various cluster parameters, using a component called CLFIT. CLFIT uses X-ray data in the form of `datacubes', which are composed of a stack of 2-D images of the cluster. The energy range within a datacube is typically around 0.5 to 7 keV, divided into 32 slices, each of which corresponds to a single image in the cube. 

The analysis is performed using a `forward fitting' method; the gas properties of the data are described by functions of cluster radii, and defined in a series of discrete spherical shells. The X-ray emission in each shell is calculated using a plasma spectral code. The plasma codes, specifically MEKAL \citep{kaa92}, and Raymond-Smith \citep{ray77}, model the emission from optically-thin, collisionally-ionised hot plasma. This is done by calculating ionisation and recombination rates, which are then used to model the continuum and line emission from the plasma. The resulting spectrum is then given a redshift correction corresponding to that of the cluster being studied. The data are then convolved with the spectral response of the appropriate X-ray detector and projected into a cube. The cube is then blurred with the detector Point Spread Function (PSF), producing a datacube which can be compared with the observed data.

This forward-fitting approach to analysis has particular advantages over other fitting techniques such as conventional deprojection. The detector PSF blurring is automatically incorporated into the datacube, and multiple datasets can be fitted simultaneously. It is also possible to extrapolate fitted models beyond the data in order to allow for the presence of projected flux from large radii. Another advantage, as discussed later, is that forward-fitting readily permits the application of Bayesian analysis methods. 

To test the goodness of fit of the model with the data, the fit is parameterised with a likelihood statistic. The Cash statistic \citep{cas79} is used in preference to a chi-squared fit because the binned data often contain only a few counts, and so are highly Poissonian. The model parameters are then modified iteratively in order to minimise the Cash statistic, thereby obtaining the best fit to the data. The standard minimisation technique used is a variation of the Bevington CURFIT algorithm \citep{bev69}. This method iteratively tests the slope of the statistic and steps towards regions in the fit space with a lower Cash statistic, until a predetermined minimum gradient is achieved.

The cluster-fitting software analyses the data in the form of a datacube, so it was necessary to convert the cleaned Chandra data into the appropriate format. First of all appropriate cube dimensions were selected, in this case the image dimensions were 512x512 pixels, with 32 energy slices up to 7.0 keV. The image and background data were generated using the CIAO dmcopy command, and the weighted arf and rmf response files were extracted using mkacisrmf. Starlink utilities were then used to construct the datacube, along with background and vignetting cubes.

The models selected for fitting with the Chandra data was a linear temperature model with a cooling flow component, known as CTLF. The cooling flow emission is incorporated directly into the model, rather than being an additive component, as it displaces the central emission. The temperature is parameterised according to the following equation:

\begin{equation}
T(r)=T_{norm}+rT_{gradient},
\end{equation}
where $T_{norm}$ and $T_{gradient}$ are respectively the temperature normalisation and gradient. The gas density is parameterised using the King profile:

\begin{equation}
\rho_{r}=\rho_{norm}*(1+(r/r_{core})^2 )^{(-\rho_{index})},
\end{equation}
where $\rho_{norm}$, $r_{core}$ and $\rho_{index}$ represent the density normalisation, density core radius and density index respectively.

The cooling flow uses three parameters; the cooling flow radius $r_{cool}$, the cooling flow density index $\rho_{cfindex}$ and the cooling flow temperature index $T_{cfindex}$. Inside the cooling radius, the model incorporates the cooling flow parameters, and the above equations are modified to:

\begin{equation}
T(r<r_{cool})=T_{cool}*(r/r_{cool})^{(T_{cfindex})}
\end{equation}
and
\begin{equation}
\rho(r<r_{cool})=\rho_{cool}*(r/r_{cool})^{(-\rho_{cfindex})}
\end{equation}
where $T_{cool}$ and $\rho_{cool}$ are defined by the temperature and density profiles outside $r_{cool}$.

In order to prevent the density and temperature from tending to infinity and zero respectively at small radii, a fixed parameter called CF$\_$MINRAD defined a cutoff radius within which the density and temperature are held at the values ($\rho_{cfminrad}$ and $T_{cfminrad}$ respectively) they take at this radius.

In addition to these eight parameters, there are two more free parameters, representing the hydrogen column density and iron abundance. Two more parameters, which determine the RA and Dec position of the model centre, remain fixed during fitting.

\begin{figure}
\centering
\includegraphics[width=8.4cm, angle=0]{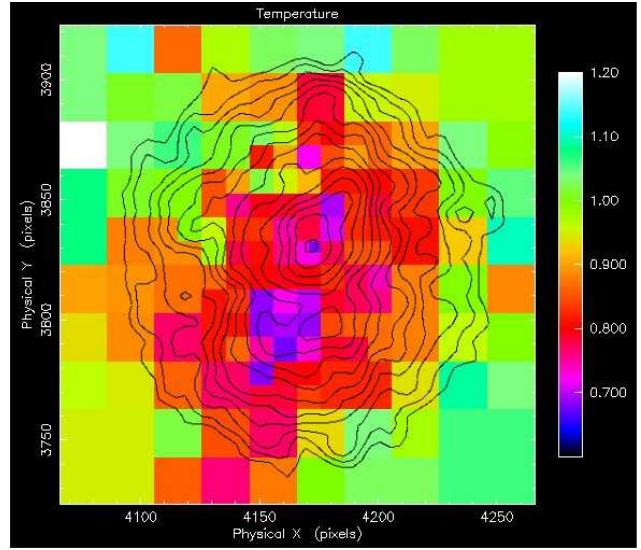}
\caption{Temperature map generated using the spectral mapping technique outlined in Section~\ref{sec:analysis_spectral}. The counts threshold for each bin was
set at 100 counts. The cool region at the core of the group is clearly discernible. Of particular interest is the apparent
northwest-southeast axial alignment, and the offset between the coolest gas and the group centre. The black outlines near the centre of the image are brightness contours from the smoothed Chandra image in Fig.~\ref{smooth}.\label{spectemp}}
\end{figure}
   
\subsection{MCMC Bayesian Analysis}
\label{sec:analysis_mcmc}

A disadvantage of the standard cluster fitting minimisation is that it is not possible to clearly visualise the model parameter space, and thereby determine whether a global minimum has been found. Also the initial parameter values for minimisation should be carefully chosen, as a gradient search is often less effective in regions of the parameter space far removed from the global minimum. One solution is to use a genetic algorithm \citep{llo00} or minimisation by simulated annealing \citep{san03} to help locate the global minimum. However, an increasingly common solution to fitting in multi-dimensional parameter spaces is to calculate a Bayesian probability distribution over the model parameter space \citep[e.g.][]{lew02, mar05}.

In order to perform data analysis via a Bayesian approach, a \emph{joint probability} is constructed over all observable data, and model parameters. The most useful quantity in Bayesian inference is the posterior probability, $P(\theta|D)$, or posterior, which represents the probability of a particular model, $\theta$, given the data, $D$:

\begin{equation}
P(\theta|D)=\frac{P(D|\theta)P(\theta)}{P(D)}
\label{Bayes}
\end{equation}

In a Bayesian approach to cluster fitting, the objective is to obtain a posterior probability for all the parameters of the model. In order for the Bayesian approach to be effective, an efficient method of mapping the posterior over the multi-dimensional parameter space is required. One attempt to solve this problem has been the development of Markov-Chain Monte Carlo (MCMC) techniques.

A Markov-Chain is a sequence of random variables ${(X_0, X_1,...)}$ with the property that the distribution of $X_{n+1}$, given all previous values of the process, $X_0, X_1...,X_{n}$ depends only upon the previous state, $X_{n}$. Ideally, the chain gradually \emph{forgets} its initial state, $X_0$, and converges to a unique stationary distribution, $P(X)$, which in the present case will be the posterior given by equation \ref{Bayes}. This \emph{converged} distribution does not depend on the starting position, $X_0$ or the total number of samples $N$.

In the cluster-fitting software, the Markov Chain is constructed using the Metropolis-Hastings algorithm \citep{met53, has70}, the transition kernel is chosen such that the Markov-Chain has a stationary asymptotic distribution equal to $P(X)$, which is the distribution being sampled from. This is done using a \emph{proposal density} distribution $q(X_n,X_{n+1})$ to propose a new point $X_{n+1}$, while the chain is at position $X_n$. The proposed new point is accepted with a probability

\begin{equation}
\alpha{(X_n,X_{n+1})}=min\left(1, \frac{P(X_{n+1})q(X_{n+1},X{_n})}{P(X{_n})q(X{_n},X_{n+1})}\right)
\end{equation}
such that the transition matrix, $\mathcal{T}$ is given by 

\begin{equation}
\mathcal{T}(X_n,X_{n+1})=\alpha(X_n,X_{n+1})q(X_n,X_{n+1}).
\end{equation}
Consequently, $P(X)$ is the equilibrium distribution of the Markov-Chain, such that

\begin{equation}
P(X_{n+1})\mathcal{T}(X_{n+1},X_n)=P(X_n)\mathcal{T}(X_n,X_{n+1}).
\end{equation}

If, after this evaluation step, the candidate point is accepted, the next state becomes $X_{n+1}$. If the candidate point is rejected, the value of the chain does not change, ie $X_{n+1} = X_n$. Once the Markov Chain has been constructed, and preliminary samples have been discarded (from an initial phase known as `burn-in'), the output of the Markov-Chain will be independent samples of the stationary distribution $P(X)$. The ability to rapidly sample the model parameter space is especially useful when applied to multi-dimensional problems such as the cluster-fitting models. The CTLF model (described in Section~\ref{sec:analysis_fitting}) typically has ten free parameters during fitting. During the MCMC analysis, non-informative (i.e. uniform) priors were applied, and a multivariate Gaussian was used for the proposal distribution. All free parameters were fitted independently, and no constraints were placed on the value each parameter could adopt during the MCMC run.

The quality of results from the MCMC algorithm is rather sensitive to the step size selected for each parameter. Within the algorithm, the size of each step along a parameter is determined by multiplying a random number, 0$\leq$x$\leq$1, by a predefined value, which serves as an upper limit on the step size. If this upper limit is too high, then the number of points accepted during the MCMC run will be low (well below 50\%), as whenever the sampler arrives at a desirable region of the parameter space the large step size will cause it to sample points outside this region, which have a lower likelihood and are likely to be rejected. Conversely, selecting an upper limit on the step size which is too small will result in a high acceptance rate (above 50\%), but the sampler is likely to be confined to only a small region of the model parameter space.

In order to optimise the MCMC sampling, it is therefore necessary to select appropriate step sizes for each parameter before sampling the full multidimensional parameter space. For the analysis outlined here, the step sizes were optimised by freezing all parameters except one, and running the MCMC sampler with a `guess' step size for that one free parameter. After a few hundred jumps, the sampler was stopped and the number of accepted points examined.  If the acceptance ratio was below 45\% then the step size was decreased, or increased if the acceptance ratio was above 55\%. If the number of points accepted lay within the range $45\%\leq n\leq 55\%$, then the step size was accepted as appropriate, the free parameter was frozen, and the next parameter was thawed and the process repeated. When all parameters have appropriately scaled step sizes, the main MCMC run was performed with all the parameters freed simultaneously.

\begin{figure}
\centering
\includegraphics[width=8.6cm, angle=0]{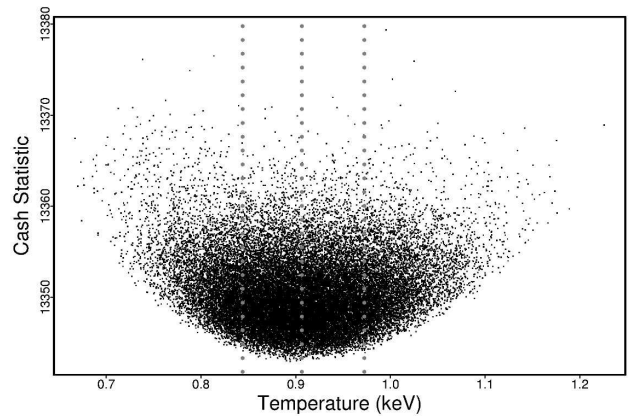}
\caption{Sample of the output from an MCMC run with the ROSAT NGC 4325 data. Each successful jump within the parameter space is represented by a point. The vertical lines indicate the best fit, upper and lower bounds found using cluster-fitting minimisation (see Section \ref{sec:analysis_fitting}). Note how the points clearly delineate the value of the Cash statistic over a broad range of parameter values, and that the number density of the points peaks around the minimum Cash statistic value. \label{mcmc}}
\end{figure}

\begin{figure*}
\centering
\includegraphics[width=12cm, angle=270]{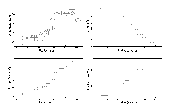}
\caption{Radial profiles for NGC 4325 obtained by performing a deprojection analysis with XSPEC, as described in Section~\ref{sec:analysis_xspec}. Bars indicate deprojection results, and the diamonds in the temperature plot show results of a projected fit to the same data.\label{temp}}
\end{figure*}

\subsection{2D Spectral Mapping}
\label{sec:analysis_spectral}

The Chandra data were prepared as outlined in Section~\ref{sec:chandra_reduction}. However, rather than applying a blank sky background, an outer annulus was selected on CCD, chip 7, centred on the group X-ray emission and outside the region selected. A new events file was created with an energy range between 0.5 and 7 keV, which was then used in an automated procedure which iteratively subdivided a specified square region. A spectrum was extracted for the entire 600x600 pixel grid used for the spectral map, using the dmextract command in CIAO, and the same was done for the background region. These spectra were then fitted to a simple MEKAL model within the Chandra fitting package, Sherpa, which produced a temperature, metallicity and chi-squared value for each pixel in the grid.

The 600x600 pixel square was then divided into four 300x300 pixel squares. If the number of counts within each of these squares was above a predefined threshold level (usually between 100 and 200 counts), then spectra and responses were obtained for each square and they were then fitted in sherpa to find temperature, metallicity and chi-squared values. The pixel values for a smaller region then replaced those of the larger region which preceded it. In this way, a spectral map is gradually built in which the size of the region is proportional to the number of counts it contains. The map is completed when no square in the map contains more counts than the predefined threshold level.

\section{Results}
\label{sec:results}

\subsection{2D Spectral Mapping}
\label{sec:results_spectral}

The results of the spectral temperature mapping are shown in Fig.~\ref{spectemp}. The spectral map revealed an interesting feature in the core of the group; an asymmetrical region of cool gas, with a temperature between 0.6 and 0.7 keV. This feature is not apparent in the smoothed X-ray image (Fig.~\ref{smooth}), although the northwest-southeast alignment of the cool gas does appear to coincide with some of the brightest X-ray features within the group centre. By comparing the temperature map to an DSS optical image of the group centre (Fig.~\ref{optical}), it appears that the coolest gas lies close to the southeastern edge of the central elliptical galaxy, with the extended cool feature extending further southeast to beyond 30kpc of the group centre.

The orientation of various features within the group appear to vary according to their distance from the group centre. The brightest X-ray surface-brightness features and the coolest gas, which are within 10kpc of the centre of the group, appear aligned northwest-southeast, with the coolest gas approximately 7kpc from the group centre. The elliptical galaxy which encompasses these features is aligned north-south. This north-south alignment is visible in the smoothed X-ray data out to a distance of approximately 20kpc from the group centre, after which the gas distribution appears to be much more spherically symmetric.

\subsection{XSPEC deprojected radial profiles}
\label{sec:results_xspec}

The temperature profile obtained from the XSPEC deprojection analysis is shown in Fig.~\ref{temp}. The deprojected data are represented by bars, and the projected data by diamonds. The profile shows a steep drop in temperature around 50 acrseconds (25 kpc) from the group centre. This distance coincides with the apparent departure of the gas distribution from spherical symmetry, as indicated in the smoothed Chandra image (Fig.~\ref{smooth}). The temperature of the gas at the centre of the group is approximately 0.7 keV, which is consistent with the temperature obtained from 2D spectral mapping.

The XSPEC analysis of the Chandra data indicated the presence of excess absorption above the expected galactic column. The expected level, for example that provided by HEASARC's nH FTool, is approximately $0.224$x$10^{21}$cm$^{-2}$. The level indicated by the XSPEC analysis, however, was much higher, at $\sim$0.7x$10^{21}$cm$^{-2}$, with an increased level of $\sim$1.1x$10^{21}$cm$^{-2}$ within 30kpc of the group centre.

An excess absorption level was also noted by Mushotzky et al., when data from an XMM observation of NGC 4325 were analysed \citep{mus03}. It is interesting to note that the level of excess absorption mentioned by Mushotzky et al. is $\sim$0.6x$10^{21}$cm$^{-2}$, although they described the excess falling gradually with radius, reaching galactic levels by $R>200^{\prime\prime}$. We also observe a decrease in the absorbing column beyond $30^{\prime\prime}$ in our XSPEC deprojection analysis, although it does not appear to fall back to the expected galactic level, despite our analysis extending to just beyond $200^{\prime\prime}$. It is interesting to note that excess absorption has also been reported in other groups and clusters \citep[e.g.][]{ewan03, pra05}, from both Chandra and XMM-Newton data, although we were unable to find an example where the cause has been sucessfully determined.

In an attempt to explain the cause of this excess, infra-red survey data was examined in the group's vicinity. The IRAS Sky Survey Atlas 100$\mu$m survey image shows some dust features in the region. To determine whether these features could be responsible for the excess absorbing column in the direction of the group, the mean intensity of the infrared emission was rescaled to match the hydrogen column density, as outlined in \citet{miv02}. However, this did not produce a significant excess in absorbing column.

Another potential origin for the excess column could be a deficit in soft X-ray emission in the direction of NGC 4325, compared to the all-sky average (and hence the blank sky background, which is assembled from a whole set of pointings). To check for this, we  
examined the ROSAT all-sky survey data, and  estimated the mean background count rate in an annulus $r=0.5^{\circ}$ - $1^{\circ}$, centred on NGC 4325. The integrated mean count rate of ($1.189\pm0.054$) counts per second in the ROSAT R45 combined bands (0.47 to 1.21 keV) did not appear deficient when compared with other groups and clusters. So, at the end of these investigations, we are unable to explain the excess absorption. However, since the hydrogen column density was left as a free parameter in both the cluster-fitting models and the XSPEC deprojection, this excess absorption will be modelled out in the fitting, and so should not bias other fitted spectral parameters.

\subsection{MCMC and minimisation}
\label{sec:results_mcmc}

\begin{figure}
\centering
\includegraphics[width=8.0cm, angle=0]{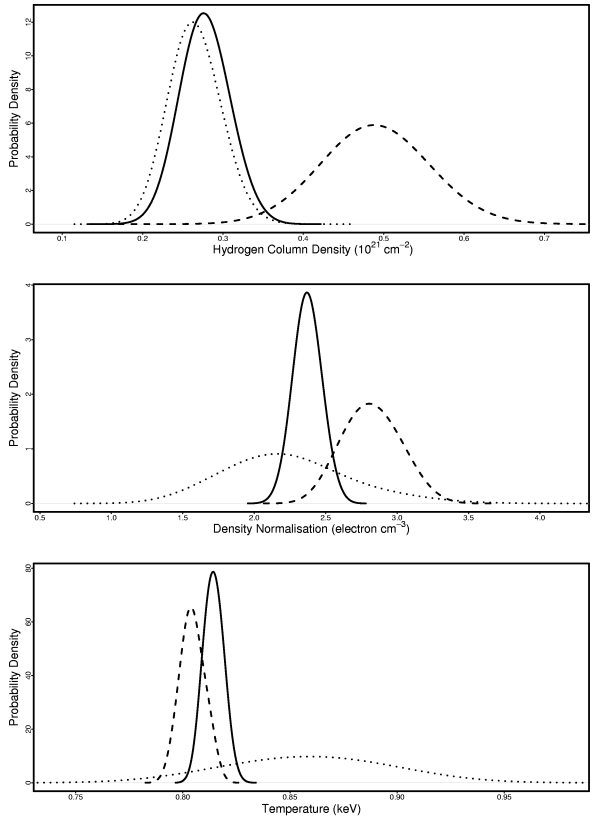}
\caption{Posterior probability distributions for three parameters at the best-fit cooling radius of $\sim$15 kpc, obtained using MCMC to sample the model parameter space. The results are shown for Chandra (dashed), ROSAT (dotted) and joint data (solid). The superior constraints of the Chandra data over the ROSAT data are clearly visible, except for the hydrogen column density. Note that in all cases the joint fit results are better constrained than those of Chandra or ROSAT individually.\label{probs}}
\end{figure}

The cluster-fitting software was used to analyse the ROSAT, Chandra and joint data, using both the CURFIT or gradient-based minimisation method and MCMC. The ROSAT datacube used with the fitting software in this analysis was originally part of the ROSAT X-ray sample in \citet{san03}. The MCMC results for the ROSAT temperature parameter are shown in Fig.~\ref{mcmc}, together with the minimum and limits found using minimisation (the vertical lines in Fig.~\ref{mcmc}). There is excellent agreement between the results obtained by the two methods; in Fig.~\ref{mcmc} the minimum Cash statistic found using both techniques have the same temperature value ($\sim$0.905 keV). Further results from the MCMC analysis are shown in Fig.~\ref{probs} and discussed in Section~\ref{sec:discuss_densities}.

One of the significant advantages of the MCMC technique is that it enables the parameter space to be easily visualised. Single-parameter plots such as that shown in Fig.~\ref{mcmc} are trivial to generate from the output data, as are two-parameter plots, which are particularly useful for finding correlations between different model parameters. This has clear advantages over the gradient minimisation output, from which it is difficult to establish whether a located minimum is global or just a local minimum. One important contribution of the MCMC analysis, therefore, was to validate the results of the gradient minimisation, by demonstrating that it did indeed find a global minimum value and had appropriately assigned upper and lower error bounds.

The MCMC technique also makes it easier to explore the full model parameter space than gradient minimisation methods, as successful results are much less dependent on the choice of initial position in the parameter space. If the initial coordinate in the parameter space is far removed from the global minimum, then gradient minimisation methods will usually struggle to find the minimum, as the Cash statistic may become insensitive to changes in some of the parameter values. The MCMC technique, however, is able to rapidly move towards appropriate regions of the parameter space, in an initial phase usually referred to as `burn in'. These initial points must usually be removed from the final MCMC data to avoid biasing the results through the inclusion of a trail of points from the initial starting position. One way to avoid the need to remove burn-in points from the data is to select an appropriate starting position, most conveniently by running the algorithm twice; once to find the minimum region of the parameter space, and then again, starting from a point within this region. 

\section{Discussion}
\label{sec:discuss}

\begin{figure}
\centering
\includegraphics[width=8.6cm, angle=0]{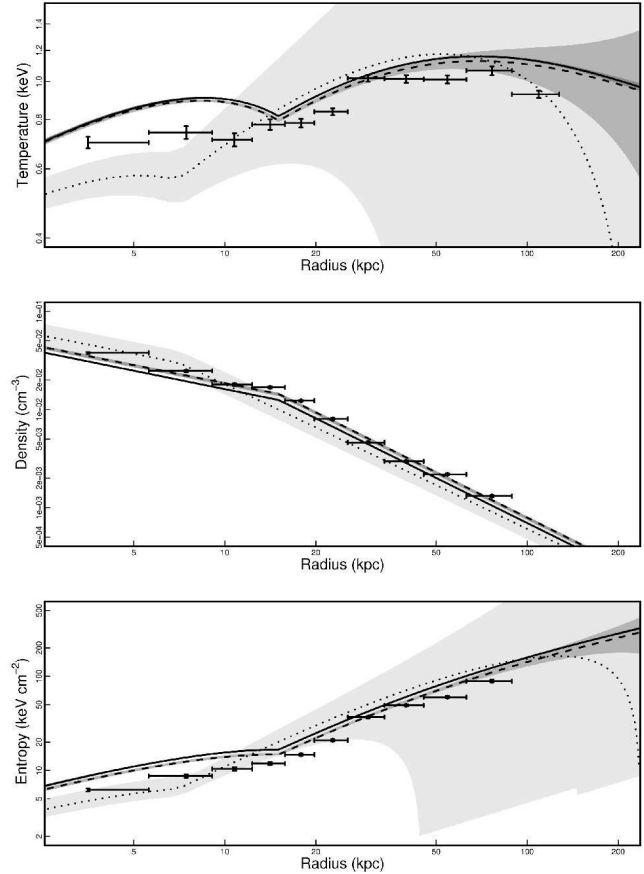}
\caption{Plots of gas parameters derived from cluster model fitting with XSPEC deprojection (points with error bars) from Chandra data. Chandra, ROSAT and joint-fit results are the dashed, dotted and solid lines respectively. Cluster-fitted data are best-fit values, with upper and lower bounds of 1 sigma confidence indicated by the grey shaded regions, with dark grey and light grey for Chandra and ROSAT respectively. The joint-fit errors are indicated by the darkest shade of grey, although at the scale drawn above, the bounds generally lie within the thickness of the solid line representing the joint best-fit.\label{profiles}}
\end{figure}

\subsection{Combining ROSAT and Chandra data}
\label{sec:discuss_combine}

A key advantage of the forward fitting approach, compared to deprojection, is that it allows datasets to be fitted jointly, with the background and vignetting of each dataset being handled separately. There are clear advantages to be gained by combining data from both Chandra and ROSAT, because the information provided by these instruments is complementary. Chandra has the advantage of high resolution imaging over a small field of view, whereas ROSAT, despite having a lower resolution, has a much larger field of view. By combining data obtained with both instruments, information about NGC 4325 extends from the group core, out to a radius of $0.1\,^{\circ}$.

The results shown in Fig.~\ref{profiles} show model parameter values for joint-fitted data, along with those obtained by fitting Chandra and ROSAT data individually. The joint fit was clearly influenced most strongly by the Chandra data, which is to be expected, as the Chandra datacube contained approximately 15 times as many counts as the ROSAT datacube. Each dataset on the plot consists of a best-fit value, within an upper and lower error `envelope'. The joint-fit and Chandra errors are small enough that only the ROSAT bounds are clearly visible at smaller radii. The plots demonstrate the good agreement between the ROSAT and Chandra data, although for the entropy and particularly temperature profiles the ROSAT error envelope is very large. The size of the ROSAT error envelope can reduced by choosing alternative models, although the chosen model was more flexible and hence able to replicate the features seen in the XSPEC temperature profile. The entropy profile is well constrained, with a break around 20kpc and a slope of $\sim$1.1, which is steeper than that typical of other observed groups \citep{sun03, mus03}. The density results indicate that a power-law profile is a good approximation for the gas density.

One way to evaluate the quality of fit provided by the forward-fitting method is to compare the radial profiles with the results obtained using the XSPEC deprojection method outlined in Section~\ref{sec:analysis_xspec}. Fig.~\ref{profiles} shows the XSPEC radial profiles along with the appropriate cluster-fitted radial profiles. There appears to be reasonable agreement between the results for the density, but with some disagreement in the core for the temperature and entropy profiles.  When fitting the Chandra data, the density profile initially appeared to be systematically higher than that obtained by the XSPEC deprojection analysis. This was resolved by masking out the cool gas located 7 kpc southeast of the group centre. Since emission from this region is not excised in the deprojection analysis, the fitted temperature and entropy profiles would be expected to deviate from the deprojection results within 10 kpc of the group core.

There is some difference between the shapes of the XSPEC and cluster-fitted temperature profiles. The deprojection results suggest an almost flat temperature profile within 10 kpc of the core, rising between 10 and 20 kpc. The fitted model does not appear to reproduce this shape. One possible explanation for this discrepancy is that the parameterisation of the cluster-fitting model is too inflexible to allow for the full variation within the temperature profile. In particular, the cooling radius, which is clearly visible as a break in the XSPEC density profile, does not have a clear corresponding feature in the XSPEC temperature profile. The cluster-fitting model is forced to link the density and temperature profiles, producing a dip in the temperature profile at the cooling radius.

Another possible explanation for a discrepancy between the XSPEC and cluster-fitted temperature profiles is spectral bias in the XSPEC deprojection results. It has been noted that there is a discrepancy between the projected spectroscopic temperature and the emission-weighted temperature, which can cause the projected temperature to be underestimated by as much as 20--30 per cent \citep{maz04, ras05}. This is a result of the spectroscopic temperature being biased towards lower temperature components, although algorithms which correct for this effect have been devised \citep{vik06}. An advantage of forward-fitting is that it is immune to this spectroscopic bias.

\subsection{MCMC parameter probability densities}
\label{sec:discuss_densities}

Analysing data from ROSAT and Chandra allowed us to compare the fitted values of the various model parameters. The minimisation within the cluster-fitting has the capability of producing confidence limits for each parameter by stepping either side of the best-fit value. However, a much more informative way to examine the parameter values is to use the MCMC method to map the parameter space. The result can be seen in Fig.~\ref{probs}. The results show the marginalised posterior distributions of three parameters. Marginalisation of parameters from the MCMC results is done simply by counting the number of samples within binned ranges of each parameter value. The distribution of binned parameter values were then converted into a density function using a kernel density estimator.

By comparing the parameter values from each dataset, the relative constraints for each parameter become readily apparent. In general the Chandra model parameters are more tightly constrained than those of the ROSAT model, as noted previously. However, the hydrogen column density appears to be an exception. This may be related to the excess galactic HI absorption discussed in Section~\ref{sec:results_xspec}. In this context, we note that the soft X-ray calibration of Chandra, which has been repeatedly improved in recent years. However, using the latest version of CIAO available (3.4) did not reduce the observed excess in the absorbing column. It is also possible that the higher sensitivity of ROSAT at lower energies, compared with Chandra, meant that ROSAT was better able to produce a reliable detection of the local N$_H$ level. The Chandra spectra used in both the XSPEC deprojection and the cluster-fitting analysis had a minimum energy of 0.5 keV, whereas the ROSAT spectra extended down to 0.2 keV.

Another potential cause is that the ROSAT data reduction was performed using a local background, whereas during the Chandra data reduction a blank-sky background was used. It is possible that this blank-sky image contained a different level of absorbing column and/or soft galactic excess, compared to that found at the location of NGC 4325. To determine if this was the case, ideally a local background on the Chandra chip should be used to compare with the blank-sky background. Unfortunately the option of using a local background was not available to us, as the source filled the primary chip and the other back-illuminated chip was not switched on during the observation. Comparing the rescaled blank-sky background file to those used when examining other groups and clusters did not indicate that the NGC 4325 blank-sky file was in any way exceptional.

During both minimisation and the MCMC mapping of the ROSAT data, it was clear that the cooling flow parameters were poorly constrained. This is unsurprising, not only because of the low resolution of the ROSAT data, but also because when the ROSAT datacube was originally constructed, the innermost arcminute of data were removed to improve fitting in the original analysis \citep{san03}. As a result, only the Chandra data could provide reasonable estimates for the cool-core parameters. Fortunately the high resolution of the Chandra data meant that these parameters were well constrained.

\subsection{Possible cavities in the X-ray surface brightness}
\label{sec:cavities}

To determine whether the best fit model from the cluster-fitting was a good representation of the data, a residual image of the model and data was produced. Firstly, a \emph{predicted} datacube was created from the fitted model, using the `CLOUT' routine within the cluster-fitting software. This routine writes the data generated from the fitted model directly to a file, rather than using it for fitting. This sample model datacube was then subtracted from the original Chandra datacube and projected along the energy axis, to produce the residual image.

\begin{figure}
\centering
\includegraphics[width=8.4cm, angle=0]{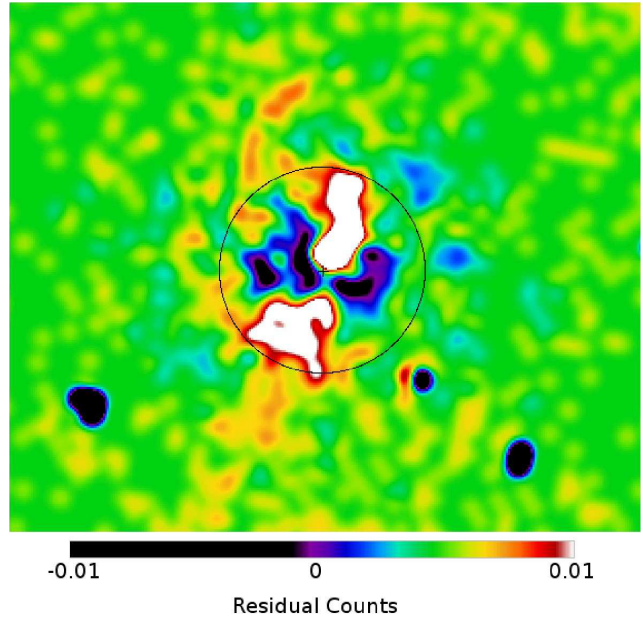}
\caption{Gaussian smoothed residual image created by subtracting model-derived data from the Chandra cluster-fitting datacube. The cross at the centre of the image indicates the centre of the subtracted CTLF model, and is enclosed by a circular overlay at 20kpc (compare with Fig.~\ref{smooth}). Note the apparent cavities either side of the group centre. The dark circular regions in the lower half of the image are removed point sources.\label{residual}}
\end{figure}

The residual image (Fig.~\ref{residual}) indicates regions of the Chandra data which depart from the cooling-flow model. This image indicates that the model generally agrees well with the Chandra data; most of the residual features lie at the centre of the group, where departures from a simple model are expected (for example from Fig.~\ref{smooth}). The most striking features are two apparent cavities east and west of the group core. In addition, there is some excess emission to the southeast, in the same location and orientation as the cool gas indicated in the spectral temperature mapping (Fig.~\ref{spectemp}). In Section~\ref{sec:agn} we examine the implications of these observed features in the context of possible AGN heating.

\begin{figure}
\centering
\includegraphics[width=8.4cm, angle=0]{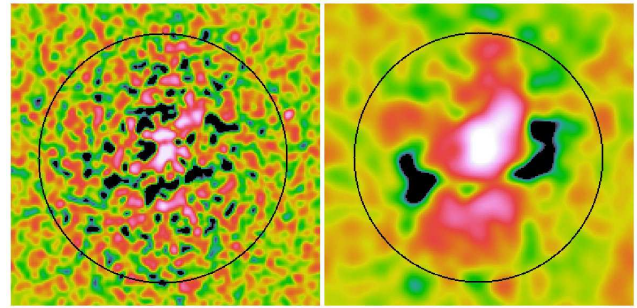}
\caption{Unsharp mask X-ray surface brightness images of the group centre, created by subtracting an image smoothed with a Gaussian of sigma 5kpc from one smoothed with a Gaussian of sigma 0.5kpc (left) and 1.25 kpc (right). The black circular overlay has radius 20kpc for comparison with Fig.~\ref{residual}. Note the presence of the dark regions indicating a deficit in X-ray surface brightness, in the same location as those seen in the residual image.\label{unsharp}}
\end{figure}

Since our model is azimuthally symmetric, there is a possibility that the bright and dark features in the residual image could be caused by ellipticity in the data, rather than actual cavities. To explore this, an unsharp mask image was generated from a cleaned Chandra X-ray image of the group. Unsharp masking is used to pick out features of a particular scale size within an image, by subtracting a highly smoothed image from a less smooth one (e.g. \citet{fab06}). The relative smoothing scales of the two images determine the size of the features which are sharpened. For the right-hand image shown in Fig.~\ref{unsharp}, the unsharp mask was generated by subtracting a highly smoothed copy of the image, using a Gaussian with a sigma of 20 pixels ($\sim$5 kpc), from another copy smoothed using a Gaussian with a sigma of 5 pixels ($\sim$1.25 kpc). Using these smoothing scales, the unsharp mask image reproduces the cavities seen in the residual image.

However, the unsharp masking employs a {\it circular} Gaussian, so smoothing on a large a scale could also generate spurious `cavities' if the surface brightness distribution is strongly elliptical. To check this, we reduced the smoothing scales by a factor of 2.5. The result, shown in the left-hand image of Fig.~\ref{unsharp}, still shows a general (albeit noisier) flux deficit to the east and west of the
cluster centre.

\begin{figure}
\centering
\includegraphics[width=8.0cm, angle=270]{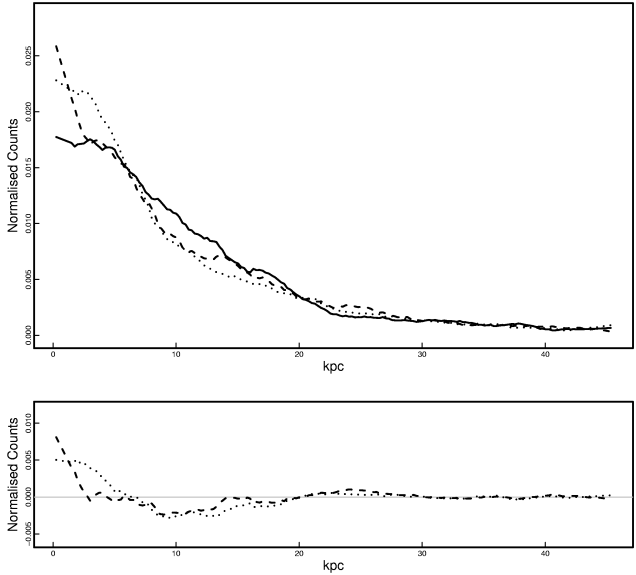}
\caption{The upper panel shows the profiles of smoothed X-ray counts to the east (dashed) and west (dotted) relative to the north (solid) of the group centre, extracted from a cleaned X-ray image of the group. The profiles have been renormalised to have identical total integrated counts. There is a clear deficit in mean photon count between 5 and 15 kpc in the east and west profiles, relative to the north. The lower panel shows the result of subtracting the northern profile from the east (dashed) and the west (dotted) profiles. \label{cuts}}
\end{figure}

As a final check, in order to suppress the noise without any circular smoothing at all, we compare profiles of surface brightness in different directions.  This was done by extracting the projected mean photon count within rectangular regions to the north, east and west in the unsmoothed Chandra X-ray image. The south was not included due to the presence of an X-ray bright region $\sim$8 kpc from the group centre. The profiles were derived from rectangular regions of 180 pixels by 60 pixels, with the longer dimension directed radially away from the group centre. The counts were then projected along the 60 pixel direction, and each extracted profile was then rescaled to contain the same total
number of counts. This was done to correct for the ellipticity in the group profile, which otherwise would have resulted in higher integrated counts to the north. The smoothed results (Fig.~\ref{cuts}) show a clear deficit in the mean photon count for the east and west samples, relative to the north, around 10kpc from the group centre, which is where the cavities are seen. It is important to note that the radial profiles all converge beyond 20 kpc, which implies that the deficits cannot be explained simply as an effect of ellipticity.

\subsection{AGN Interaction and the Cool Core}
\label{sec:agn}
The smoothed residual image (Fig.~\ref{residual}) and the unsharp mask image (Fig.~\ref{unsharp}) reveal the presence of a pair of cavities in the intragroup medium (IGM) resembling those seen in clusters with ongoing interaction from a central active nucleus \citep[e.g.][]{bir04}. However, unlike most cavities discovered previously, these features are not easily visible in the raw X-ray images. This reflects their unusually small size ($\sim5$ kpc) which results in the fractional line-of-sight deficit in surface brightness caused by the cavities being lower than for larger cavities, so that there is a small fractional contrast between the cavities and the bright core. There is no detectable radio emission associated with these cavities; no radio sources were detected in the FIRST survey, down to 0.99 mJy/beam, and none were found in the VLA all-sky survey (NVSS) within 1 arcmin of the central X-ray emission.

While the presence of cavities might indicate a recent outburst, the lack of corresponding radio emission suggests that enough time must have elapsed to allow the radio lobes to fade. That some time has passed since AGN heating last occurred is also implied by the steep entropy profile, which only shows slight flattening within 20 kpc of the group centre. This is in contrast to the two clusters analysed by \citet{don05}, which host cool cores with no signs of central radio activity or cavities in the IGM. These clusters have elevated entropy profiles, with flattening observed to much larger radii than is seen in NGC 4325. This suggests that rather than being subject to an exceptionally large outburst sometime in the past, NGC 4325 has been able to cool rapidly after a small AGN outburst. The displacement of gas in the group centre by small cavities would also explain the flattening the entropy profile within r $\sim$15 kpc.

Given the very short cooling time of the gas in the core ($\sim$0.05 Gyr), it seems likely that NGC 4325 is currently in a pre-outburst stage, in which rapid cooling has radiated away much of the energy injected from a previous outburst. Continued rapid cooling at the present level could quickly (i.e. within $\sim$ 50 Myr, implied by the cooling time) lead to sufficient cooling of gas to trigger the next outburst.

It is possible to estimate the work done by the bubbles in heating the IGM \citep[e.g.][]{all06,bir04}, using a characteristic timescale, $t_{c}$ which can be defined such that:

\begin{equation}
t_{c}=\frac{R}{c_{s}}
\end{equation}
where R is the distance from the centre of the bubble to the AGN, and $c_{s}$ is the sound speed. Assuming that the bubble has expanded sub-sonically, the sound speed defines an upper limit on the time taken for the bubble to expand and rise. This timescale can then be used to infer a lower limit on the rate of work done on the IGM by the bubble. Assuming the bubble centres are 10 kpc from the group centre, with a gas temperature of $(0.69\pm0.02)$ keV, we obtain $t_{c}=(2.25\pm0.03)\times10^{7}$ years.

As there is no evidence of AGN activity from radio observations, we may assume that relativistic particles are no longer being injected into the bubbles. Additionally, as the bubbles are still close to the core, it appears that they have not yet begun rising buoyantly. Assuming that the bubbles are currently in pressure equilibrium with the surrounding IGM, and that they are supported by a relativistic fluid, we can estimate their internal energy, $E_{b}$:

\begin{equation}
E_{b}=3PV_{b}
\end{equation}
where $V_{b}$ is the bubble volume and $P$ is the pressure of the IGM. At 10 kpc from the group centre, the pressure obtained by our deprojection analysis was $P=(4.23\pm0.17)\times10^{-11}$ erg cm$^{-3}$, from which we estimate $E_{b}$ to be approximately $(1.24\pm0.05)\times10^{57}$ erg. The \emph{total} work done on the IGM by the bubbles can then be calculated by adding together the work done by the bubble as it undergoes adiabatic expansion and the work done on the IGM when the bubble is initially created (A. Babul, private communication):

\begin{equation}
E_{IGM}=3PV_{b}(x^{1/4}-1)+PV_{b}x^{-3/4}
\end{equation}
where $x=P_{i}/P$, $P_{i}$ is the internal pressure of the bubble. Assuming gentle inflation by a weak AGN outburst, with $x=2$, we obtain $E_{IGM}=(4.8\pm0.2)\times10^{56}$ erg.

Using this formation energy and the timescale $t_{c}$ calculated earlier, it is possible to estimate the average rate of work done on the IGM by the bubble, given by:

\begin{equation}
W=\frac{E_{IGM}}{t_{c}}
\end{equation}
Using the results obtained previously, we obtain $W=(6.77\pm0.27)\times10^{41}$ erg s$^{-1}$ for each bubble. The unabsorbed X-ray luminosity within 15 kpc of the group centre is $L_{x}=1.5\times10^{42}$ ergs s$^{-1}$, so according to our estimate the average work done on the IGM by the bubbles is at best barely sufficient to completely counteract the cooling of gas in the group core, particularly as this value represents an upper estimate of the rate of work done. This may explain why we observe indications of an impending AGN outburst, despite seeing X-ray cavities so close to the group core.

\section{Conclusions}
\label{sec:conclusion}

The analysis of the NGC 4325 group has provided the opportunity to compare the relative merits of different analysis methods, specifically forward fitting and deprojection. The forward fitting technique described in this paper has a number of advantages over deprojection, some which have already been mentioned. These include the ability to correctly handle PSF blurring, simultaneously fit multiple datasets, and extrapolate fitted models beyond the available data. In addition, another advantage is the ease with which bad or missing data can be excluded from fitting, whereas performing the necessary volume corrections are difficult to achieve using deprojection.

There are however some deficiencies in the forward fitting approach. One issue is that forward fitting relies heavily on the availability of appropriate models. Fitting with an inappropriate model is likely to lead to biased results, and can be a particular problem if the range of available models is limited. The ability of the cluster-fitting software to test goodness-of-fit is also limited, although the Bayesian approach using MCMC provides a suitable alternative method. Rather than attempt to determine which approach to cluster analysis is superior, it would seem more appropriate to consider how they may be used together in order to take advantage of the merits afforded by each. For example, deprojection analysis can be used to derive non-parametric models, which can then motivate the selection of appropriate models for forward fitting. 

The analysis of this group shows that some properties of NGC 4325 are similar to other groups. In particular, the temperature profile (Fig.~\ref{profiles}) is in agreement with that of \citet{sun03}, who compare the properties of a sample of four 1 keV galaxy groups. Other properties, however, are not typical of other groups. This is most clearly seen when comparing the entropy profile of NGC 4325 with that of other groups, as can be seen in \citet{mus03} and \citet{sun03}. The entropy profile of NGC 4325 is much steeper than that of the other groups in their samples, and hence has a lower entropy in the core. In addition, our deprojection analysis suggests that there is only a slight flattening of the entropy profile within 20 kpc (see Fig.~\ref{profiles}) of the group centre.

Another distinctive features of NGC 4325 is the cool gas revealed by the spectral temperature map. The deprojection analysis suggests that the cooling time within 20 kpc of the group centre is less than 0.2Gyr. Such a short cooling time may encourage thermal instability, which would help to explain the temperature structure seen in the spectral temperature map. As mentioned in Section~\ref{sec:results_spectral}, the largest region of cool gas is offset from the group centre by $\sim$7kpc. The direction of this displacement is axially aligned with smaller blobs of cool gas closer to the group centre, as well as with the overall distribution of cool gas surrounding the innermost 25 kpc of the group. This displacement suggests that the gas may be sloshing around the group centre \citep[e.g.][]{mar01}, along the observed northwest-southeast axis.

There are at least two possible mechanisms by which the cool gas could become displaced from the group centre: merging activity or an AGN outburst \citep[e.g.][]{maz104}. A merger would likely have displaced the dark matter halo of the group, disrupting the intergalactic gas. If the cool core survived the merging activity, then such a merger could have taken place within the last $\sim$10$^{9}$ years. A slight displacement ($\sim$3 kpc) of the central elliptical galaxy from the centre of the group emission (Fig.~\ref{smooth}) may indicate that a merger has taken place and that the elliptical galaxy is still sloshing around within the core region of the group.

Alternatively, the cool gas could have been displaced by an outburst from an AGN at the group centre. With an approximate group crossing time of $\sim$10$^{9}$ years, and an inter-outburst period for the AGN of $\sim$10$^{8}$ years, the cool gas may still be sloshing around the group centre as a result of a previous AGN outburst, at a time when another outburst is about to occur. The cavities observed in the X-ray surface brightness provide evidence for past AGN activity, and estimates of the age of the bubbles indicate that they are a few $\times10^{7}$ years old. Estimating the work done on the IGM by the bubble implies that the work done by the bubbles is, at best, barely sufficient to counteract the energy lost from the IGM via radiative cooling (see Section~\ref{sec:cavities}).

Hydrodynamical simulations of the interaction between AGN and cooling flows \citep{omm03} suggest that the intracluster gas will become highly centrally concentrated just prior to another AGN outburst. As NGC 4325 has a remarkably steep X-ray surface brightness profile, this may be a good indicator that another AGN outburst is about to take place, as is the very steep entropy profile.

Given the observational evidence, we may propose a possible mechanism to explain these features. A weak AGN outburst between 10$^{7}$ and 10$^{8}$ years ago injected energy in the IGM, which generated the cavities we observe in the X-ray surface brightness. As the AGN outburst was weak, the bubbles are still only small in size (with radius $\sim$5 kpc) and difficult to see in the raw X-ray data. The weak energy of the outburst also provides a possible explanation for the lack of observed radio emission. Either the emission was so weak that it is already below the detectable threshold of the NVSS survey, or perhaps the radio emission is present, but at lower frequencies.

During the formation of the bubbles, they displaced cold gas from the core of the group, which is now seen settling back down to the centre. The largest region of cold gas, at a distance of 7 kpc southeast of the group centre, lies between the two bubbles, and may have been created by gas displaced by both bubbles as they expanded. Another consequence of the weak outburst is that its effects were largely confined to regions close to the group centre, with the expanding bubbles barely able to offset further cooling by displacing cool gas from the core. This means that the conditions required for the next outburst are already apparent, with the inner 20 kpc of the group at a temperature of $\sim$0.7 keV, and the entropy profile showing a steep gradient with only slight flattening evident close to the group centre.

There are two possible models of AGN behaviour which could explain the current pre-outburst condition of the group. The evidence described above suggests that the previous outburst was sufficiently weak that only gas within 20 kpc has been affected. Either the weak outburst was {\it typical} behaviour for this AGN, or it was for some unknown reason exceptionally weak. A succession of frequent, small outbursts from the AGN would explain why only the inner 20 kpc of the group appears to be disturbed, and why the entropy profile outside the core is unusually steep, as the gas beyond 20 kpc has been left to cool with little or no heating. Alternatively, the AGN may ordinarily be more powerful than the evidence from the last outburst suggests. If the previous outburst were unusually weak, then a much more powerful outburst may be imminent, with enough time passing since the previous larger outbursts for most of the group gas to cool and create the steep entropy profile which is observed.

One possible way to discriminate between these two scenarios would be to perform a sensitive low-frequency radio observation of the group. If a massive AGN outburst had occurred sometime in the last $\sim$10$^{8}$ years, then it may be possible to detect the tenuous remnants of former radio lobes, perhaps as far as 100 kpc or more from the group centre. This would indicate that the AGN outbursts are highly variable in power, and demonstrate that the heating effects of the AGN are not confined exclusively to the group core.

\section*{Acknowledgments}

PAR thanks PPARC for a research studentship, Nazira Jetha for useful discussions, and Arif Babul for insights regarding AGN heating. This work made use of the NASA/IPAC Extragalactic Database (NED) and the Digital Sky Survey (DSS).

\bibliography{par_bibtex}

\label{lastpage}

\end{document}